# Valley polarization dynamics of photoinjected carriers at the band edge in room-temperature silicon studied by terahertz polarimetry

Ami M. Shirai*, Yuta Murotani, Tomohiro Fujimoto, Natsuki Kanda‡, Jun Yoshinobu, and Ryusuke Matsunaga†

*The Institute for Solid State Physics, The University of Tokyo, Kashiwa, Chiba 277-8581, Japan*

*shirai-ami@issp.u-tokyo.ac.jp

†matsunaga@issp.u-tokyo.ac.jp

‡ Present address: RIKEN Center for Advanced Photonics, RIKEN, 2-1 Hirosawa, Wako, Saitama 351-0198, Japan

**Abstract**

Sixfold-degenerate valleys in Si have attracted considerable attention for valleytronics application. Using optical pump-terahertz (THz) probe spectroscopy, we study the dynamics of valley polarization in bulk Si(001) at room temperature. Linearly polarized pump pulses excite electrons and holes with asymmetric distributions in momentum space, leading to in-plane anisotropic conductivity. By varying the polarization directions of the pump light relative to the in-plane crystalline axes, the valley polarization of electrons and the momentum asymmetry of holes are separately probed by THz pulses through light-induced linear birefringence and dichroism. We demonstrate that the valley relaxation time of electrons near the conduction band minimum exceeds 1.5 ps at room temperature, in good agreement with theoretically calculated intervalley phonon scattering with *f* process. This work paves the way for Si-based room-temperature valleytronics.

Semiconductors that host degenerate valleys at the conduction band (CB) bottom or valence band (VB) top at different locations in momentum space have garnered attention for novel electronic device applications that exploit the valley degree of freedom [1-3]. While extensive studies on valleytronics have focused on two-dimensional (2D) honeycomb materials over the past decade [4-8], valley-dependent transport has been investigated since the 1970s in the indirect-gap semiconductor Si, specifically in the inversion layer of metal-oxide-semiconductor field-effect transistors (MOSFETs) [1,9,10]. Si has sixfold valley degeneracy along Δ line in the CB. To enhance carrier mobility in Si MOSFETs, it is essential to lift the valley degeneracy through strain, which suppresses intervalley scattering [11,12]. The valley degeneracy and splitting in 2D quantum wells are also related to many-body physics and spin-based quantum computation [13,14]. The generation and controllability of valley polarization have also been investigated in AlAs [15,16] and diamond [17,18], materials with similar band structures.

For valley-related physics and applications, the relaxation time of valley polarization is a key parameter that must be determined experimentally. The valley polarization in Si can be generated by interband excitation with linearly polarized light due to differences in phonon-mediated transition rates for valleys that are parallel and perpendicular to the light polarization [19,20]. Historically, valley polarization dynamics in Si has been studied using cyclotron resonance spectroscopy [20,21], which is sensitive to differences in effective masses in highly asymmetric band dispersion [22]. However, the cyclotron resonance spectroscopy for Si has mainly been employed to study excitons at low temperatures with limited time resolution. For much faster time scales, recent time- and angle-resolved photoemission spectroscopy has enabled direct visualization of carrier dynamics in momentum space with resolutions of several tens of fs [23-25]. However, intervalley scattering dynamics has been observed only for hot electrons with large excess energies (> a few hundred meV) from the valley bottom. Similarly, ultrafast carrier dynamics in Si has been studied in coherent phonon spectroscopy [26-29], extreme ultraviolet spectroscopy [30], and infrared spectroscopy [31], but these studies have also concentrated on hot-carrier dynamics. To clarify valley-related transport for potential applications, observation of valley dynamics for carriers near the band edge at room temperature is required. This has evaded experimental clarification thus far, despite considerable interest in theoretical studies [32-36].

To reveal the carrier dynamics in Si, terahertz (THz) time-domain spectroscopy combined with optical pump pulses is a promising approach, as it can detect transient conductivity close to the DC limit with sub-100 fs time resolution [37-45]. When an electron population imbalance is created between the valleys along the [100] and [010] axes by optical pump pulses, the effective-mass anisotropy for each valley results in a transient anisotropic conductivity $\Delta\sigma \equiv \sigma_{[100]} -$ $\sigma_{[010]}$, which can be probed through polarization rotation and ellipticity of THz pulses. Recently, the accuracy of THz polarimetry combined with the pump-probe scheme has improved to better than 10 μrad [46], enabling sensitive detection of anisotropic conductivity, even for the small density of photoexcited carriers near the indirect band edges. Direct observation of anisotropic conductivity dynamics in Si will provide deeper insight into valley-related physics in multivalley semiconductors, offering potential for industrial applications.

In this study, we investigate valley polarization dynamics in bulk Si(001) at room temperature using near-infrared (NIR) pump-THz probe spectroscopy. The NIR pump pulses inject carriers into the band edge with asymmetric momentum distribution, and the anisotropic conductivity is probed by THz pulses. By selecting the polarization directions of the pump pulses relative to the in-plane crystalline axes, the contributions of CB valley polarization of electrons and VB momentum asymmetry of holes near the zone center can be separately probed. The effects of nonlinear photocurrent generation and phase shifts are carefully eliminated through 2D Fourier analysis. Our results reveal that the valley relaxation time exceeds 1.5 ps at room temperature, consistent with theoretical calculations of intervalley scattering.

Figure 1(a) illustrates the experimental setup. NIR pump pulses were generated by an optical parametric amplifier and second harmonic generation with a tunable wavelength between 900 and 1100 nm and a pulse duration between 200 and 250 fs. THz probe pulses were linearly polarized in the horizontal $(x)$ direction before the sample. The transmitted THz pulses were detected by electro-optic (EO) sampling in a GaP(110) crystal using gate pulses with a 1030-nm wavelength and a 100-fs pulse duration. Two wire-grid polarizers (WGP1 and WGP2) were inserted into the THz beam path before EO sampling. When both WGPs were aligned parallel (upper configuration in Fig. 1(a)), the vertical $(y)$ component of the THz field $E_y$ was measured by EO sampling. In contrast, when only WGP1 was rotated by 45° (lower configuration), the EO sampling measured $(E_x + E_y)/2$. Thus, the two WGP configurations allowed the acquisition of both $E_x$ and $E_y$. The beam path of gate pulses was fixed, and the probe and pump time delays to gate pulses, denoted as $t_1$ and $t_2$, respectively, were independently scanned using mechanical stages. Additional details are provided in Supplemental Material [47].

The sample was a 30-μm-thick, (001)-oriented undoped Si. All experiments were conducted at room temperature. Excitation by linearly polarized pump pulses for normal incidence breaks the fourfold symmetry of the (001) plane, resulting in in-plane anisotropic conductivity depending on the orientation of the pump polarization. This anisotropy was detected as the $y$ component of the transmitted THz pulses due to linear birefringence and dichroism [63,64]. The pump and probe polarization directions used in this work are shown in Fig. 1(b). The incident THz pulses were always $x$ polarized, while the pump polarization angle was switched between -45° and +45° from the $x$ direction. Additionally, two configurations for the in-plane crystal axis were switched by rotating the sample to discriminate the contributions of the CB and VB. This is because not only the valley polarization of electrons, but also the asymmetric momentum distribution of holes is also induced by the linearly polarized pump as studied by coherent phonon spectroscopy [26-29,65-67]. In Configuration 1 (left panel) in Fig. 1(b), the (001)-oriented sample was mounted such that the $[100]$ axis was oriented by -45° to the $x$ direction in the $xy$ plane. The pump polarization along $[100]$ induces anisotropic conductivity $\Delta\sigma = \sigma_{[100]} - \sigma_{[010]}$ due to the difference in electron populations between the [100] and [010] valleys. In the $xy$ coordinate system, the conductivity tensor has off-diagonal symmetric elements $\sigma_{xy} = \sigma_{yx} = \Delta\sigma/2$ [47], which induce the polarization rotation and ellipticity of THz probe pulse. We define this pump-induced polarization component perpendicular to the incident THz pulse, $\Delta E_y$, as half the difference between the data measured with pump polarization at -45° ([100]) and +45° ([010]) to remove possible artifacts. Thus, the anisotropic conductivity dynamics is obtained by measuring the $y$ component of the probe field [47]. In addition, to observe the contribution of the VB to the anisotropic conductivity, we used Configuration 2 (right) in Fig. 1(b), where the sample was set so that the pump light was polarized along the $[110]$ or $[\bar{1}10]$ directions. In this case, the in-plane valley polarization does not occur and therefore parabolic CB does not contribute to the anisotropic conductivity $\Delta\sigma = \sigma_{[110]} - \sigma_{[\bar{1}10]}$. However, the VB momentum asymmetry of holes induces the anisotropic conductivity due to the hole effective mass asymmetry, especially for the heavy-hole band [68]. From symmetry considerations, VB asymmetric distribution might also occur in Configuration 1. However, according to previous coherent phonon spectroscopies [26-28], the VB asymmetric distribution is much stronger in Configuration 2. Thus, we assume that CB valley polarization (VB asymmetric distribution) can be probed in Configuration 1 (2), as justified by the following experimental results.

Before discussing the off-diagonal elements, we first address the longitudinal response of photoexcited carriers. Figure 2(a) presents a 2D plot of the detected THz field, $E_x(t_1, t_2)$, obtained using a 900 nm pump wavelength. The THz waveforms $E_x(t_1)$ at fixed pump delays $t_2 = -0.5, 1.0, 3.0$ ps are shown in Fig. 2(b). Figure 2(c) displays a vertical profile of $E_x$ as a function of pump delay $t_2$, where the probe delay was fixed at $t_1 = 0$ ps. The pump pulse enters the sample at $t_2 = 0$ ps, altering the probe THz waveform due to the longitudinal response of the photoexcited carriers [37]. The carrier lifetime was significantly longer than the time window of our experiment (~100 ps) due to the indirect transition in Si. From the analysis of longitudinal optical conductivity spectra $\sigma_{xx}(\omega)$ (See Supplemental Material [47]), the carrier density was evaluated as, *e.g.*, 4.6 $\times 10^{16}$ and 1.7 $\times 10^{15}$ cm$^{-3}$ for the shortest (900 nm) and longest (1100 nm) pump wavelengths, respectively. The pump-probe signal remains linear to the pump fluence [47]. Interestingly, the spectral analysis shows the presence of the Lorentzian peak at 2.5 THz on top of the Drude response only for the pump wavelength of 1100 nm [47], suggesting a partial formation of excitons when carriers are photoinjected into the band edge. The result implies remarkably suppressed interaction between phonons and cold carriers at the band edge due to limited scattering channels. In Supplemental Material, we confirmed that the THz signal is mostly determined by the Drude response and therefore the excitonic effect is negligible in the following discussion [47].

Next, we discuss the change in polarization state of the THz probe field. Figure 2(d) shows the dynamics of the peak of light-induced $y$ component $\Delta E_y$ as a function of pump delay $t_2$ up to 50 ps for a 900-nm pump wavelength. Immediately after the pump, large polarization rotation angles of up to 5.8 and 6.6 mrad were observed for Configurations 1 and 2, which rapidly decayed within a few ps. This fast relaxation of the transverse response contrasts with the long-lasting longitudinal response shown in Fig. 2(c). In addition, after the fast decay, a slower component of approximately 0.48 mrad was also discerned.

To discuss the fast relaxation signal more closely, the 2D time-domain data $\Delta E_y(t_1, t_2)$ for the 900 nm pump with 0<$t_2$<4 ps in Configuration 1 are shown in Fig. 2(e). The simplest approach to analyzing the dynamics is to cut out the vertical profile, namely $\Delta E_y$ as a function of the pump delay $t_2$ with a fixed $t_1$, as shown in Fig. 2(d). However, the 2D plot in Fig. 2(e) indicates that (i) a diagonally oscillating wavy signal appears in the ultrafast regime at $t_2 \sim 0$ ps, and (ii) the vertical signals decaying along $t_2$ are slightly tilted, suggesting a phase shift in the THz waveform $\Delta E_y(t_1)$. The (almost) vertical signal changing along $t_2$ is attributed to the anisotropic conductivity of photoexcited carriers in response to the THz probe pulse. By contrast, the diagonally oscillating signal can be attributed to nonlinear photocurrent generation during the temporal overlap of the NIR pump pulse and THz probe field. We refer to this as the field-induced photogalvanic effect, as this process can be regarded as a nominally second-order photocurrent generation by NIR pulses under inversion symmetry breaking by the THz field. A similar result has also been recently observed in THz Faraday spectroscopy for GaAs excited by circularly polarized pump pulses [61]. Although both signals of the anisotropic conductivity and field-induced photogalvanic effect overlap in the time domain, they can be separated in 2D Fourier analysis because of their different dependencies on $t_1$ and $t_2$ (See Supplemental Material [47]). Figure 2(f) shows the extracted anisotropic conductivity component $\Delta E_y^{\mathrm{ANI}}(t_1, t_2)$ in the 2D time domain. In addition, the phase of the THz waveform $\Delta E_y^{\mathrm{ANI}}(t_1)$ gradually shifts as the pump delay $t_2$ passes in Fig. 2(f), which can be attributed to slow thermalization of photoexcited carriers. To eliminate this phase shift and evaluate the information of electron population, we performed the Fourier transformation at each pump delay $t_2$. Figure 2(g) shows $|\Delta E_y^{\mathrm{ANI}}(\omega_1, t_2)|$, which represents the time evolution of the anisotropic conductivity signal in the spectral domain. Thus, we are ready to discuss the relaxation dynamics of CB valley polarization. Similarly, the same analysis was conducted for Configuration 2 to discuss the VB momentum asymmetry dynamics.

Figures 3(a) and 3(b) show the time evolutions of the spectral peak $|\Delta E_y^{\mathrm{ANI}}(\omega_1/2\pi = 1.4\ \mathrm{THz})|$ in Configurations 1 and 2, corresponding to the CB valley polarization and VB momentum asymmetry, respectively, for various pump wavelengths. The data were fitted using a single exponential function with an offset (See Supplemental Material [47]) to evaluate the magnitude and decay time of anisotropic conductivity. Figure 3(c) shows the pump photon energy dependence of the signal magnitude for Configurations 1 (CB) and 2 (VB) on the left axis. Additionally, the photoexcited carrier density $N_{\mathrm{carr}}$, evaluated from the Drude model fitting in $\sigma_{xx}(\omega)$, is also plotted on the right axis. $N_{\mathrm{carr}}$ increases monotonically as the pump photon energy rises due to the increasing absorption. While the VB momentum asymmetry signal follows a similar trend, the CB valley polarization reaches a maximum around 1.25 eV and then decreases for higher photon energies. According to the band dispersion of electrons in Si [69], the energy band $\mathrm{X}_1$ at the X-point, located just above the CB bottom, the Δ-valley, is involved in optical transitions above 1.25 eV. The result in Fig. 3(c) suggests that the proximity of this band may hinder the optical injection of valley polarization.

Figure 3(d) compares the relaxation times $\tau$ for the CB valley polarization (Configuration 1) and VB momentum asymmetry (Configuration 2) as a function of the pump photon energy. The results show distinct relaxation times for Configurations 1 and 2, confirming that the CB and VB contributions were separately observed in our polarization-selective scheme. For any pump photon energy, the CB valley polarization persists for a longer time than the VB momentum asymmetry. We found that the valley relaxation time at the band edge exceeds 1.5 ps even at room temperature, which is 20 times longer than the valley relaxation time in monolayer $MoS_2$ at room temperature [70]. In previous time-resolved spectroscopies for Si, the electron-phonon scattering of 260 fs for a pump photon energy of 1.55 eV [26] and the L-X valley scattering of 180 fs for two-photon (3.4 eV) excitation [23] were reported. Our work demonstrates that the valley relaxation time can be significantly extended at the band edge due to fewer scattering channels.

Figure 3(e) shows the relaxation rate $\Gamma = 1/\tau$ as a function of the excess energy above the bandgap, compared with the calculated intervalley scattering rate previously reported [34]. The results show a good agreement below the excess energy of ~60 meV. In contrast, the calculation exhibits a significant increase above this threshold. This behavior can be explained by considering the differences in the phonon modes involved. As schematically shown in Fig. 3(g), intervalley scattering is classified into *g* and *f* processes: the *g* process involves scattering to the opposite valley on the same axis, mediated by LO phonons, whereas the *f* process involves scattering to one of the four valleys in other directions, mediated by LA, TO, and TA phonons [71]. The calculation of the intervalley scattering in Fig. 3(e) includes both *g* and *f* processes ($\Gamma_g + \Gamma_f$) [34], whereas our experiment observes the valley relaxation rate corresponding only to the *f* process ($\Gamma_f$). The dotted line in Fig. 3(e) represents the LO phonon energy, which is approximately 60 meV [62] as shown in Fig. 3(f). The threshold-like increase in the calculation coincides with the LO phonon energy, above which the *g* process becomes dominant. Therefore, the results of this work experimentally demonstrate that the intervalley scattering rate at the band edge is dominated by the *f* process, leading to a long valley relaxation time >1.5 ps at room temperature.

In conclusion, we experimentally investigated the anisotropic conductivity dynamics of carriers photoinjected at the band edge to study CB valley polarization of electrons and VB momentum asymmetry of holes using time- and polarization-resolved THz spectroscopy. Our results demonstrate that the *f*-process intervalley scattering time for band-edge carriers exceeds 1.5 ps at room temperature, which is consistent with theoretical calculations and far beyond the previous valleytronics materials. The remarkable robustness of valley degree of freedom in Si will stimulate further investigation for their controllability and readability toward Si-based valleytronics at room temperature. Furthermore, the ultrafast, sensitive, and noncontact detection of anisotropic conductivity based on THz polarimetry techniques can also be applied in a variety of materials, which will open a new avenue for multiband semiconductor valleytronics.

## Acknowledgements

The authors thank C. Kim and H. Akiyama for their support in preparing the sample. This work was supported by JST FOREST (Grant No. JPMJFR2240), JSPS KAKENHI (Grant No. JP24K00550 and No. JP24K16988), JST PRESTO (Grant No. JPMJPR2006), and JST CREST (Grant No. JPMJCR20R4). R.M. acknowledges partial support from MEXT Quantum Leap Flagship Program (MEXT Q-LEAP, Grant No. JPMXS0118068681). R.M. conceived the project. A.M.S. and Y.M. developed the pump-probe spectroscopy system with the help of T.F., N.K., J.Y. and R.M. A.M.S. performed the pump-probe experiment and analyzed the data with help of T.F. and Y.M. All the authors discussed the results. A.M.S. and R.M. wrote the manuscript with substantial feedback from Y.M. and all the coauthors.

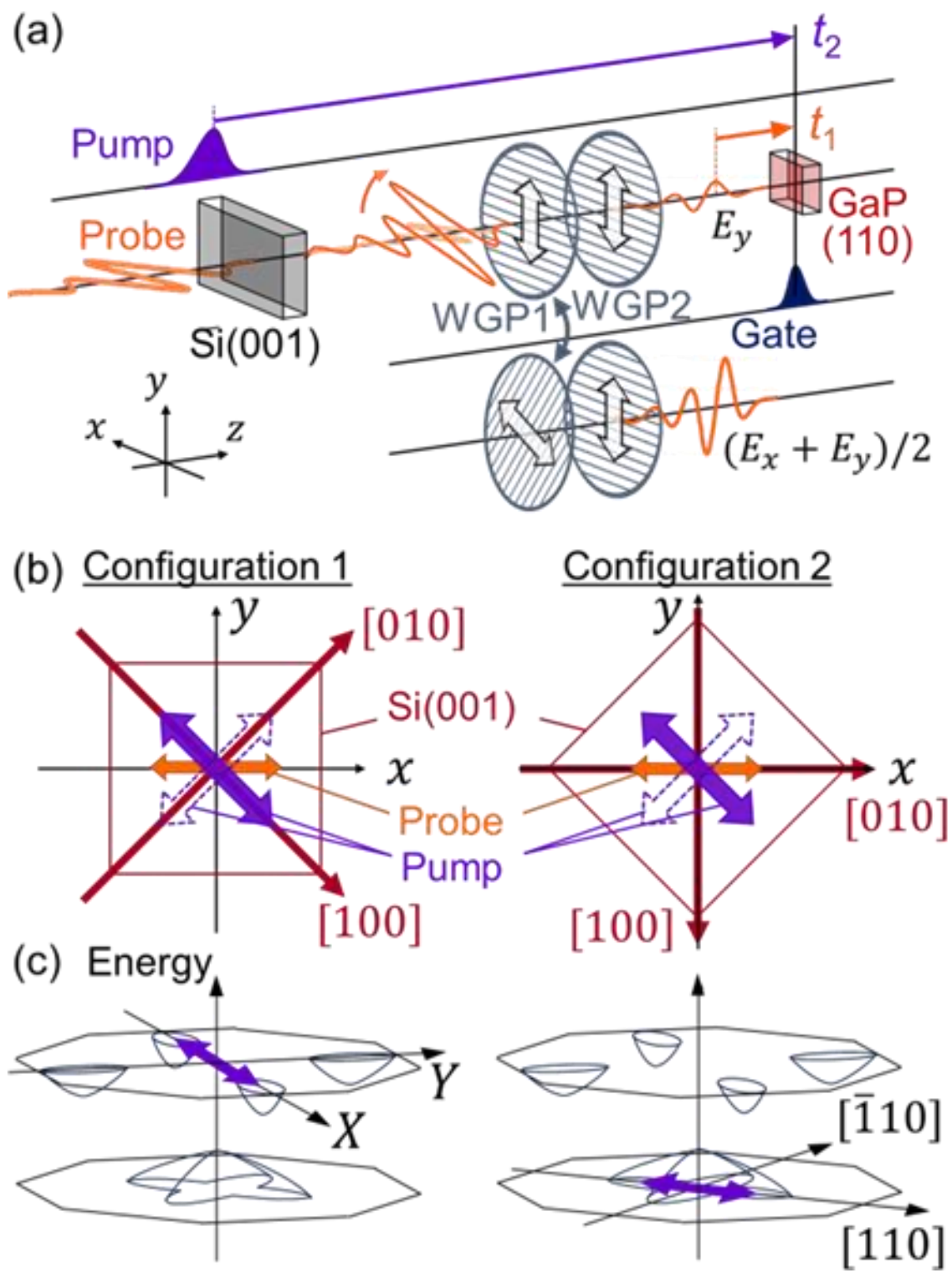

FIG. 1. (a) Schematic of the experiment using the NIR pump (purple), THz probe (orange), and gate (black) pulses. WGP: wire-grid polarizer. WGP1 was rotated to observe $E_y$ and $(E_x + E_y)/2$ separately (see text). (b) Configurations of the pump (purple) and probe (orange) polarization directions relative to the crystalline axes (dark red). The pump direction alternates between the solid and dotted arrows. (c) Schematics illustrating CB valley polarization and VB momentum asymmetry in momentum space.

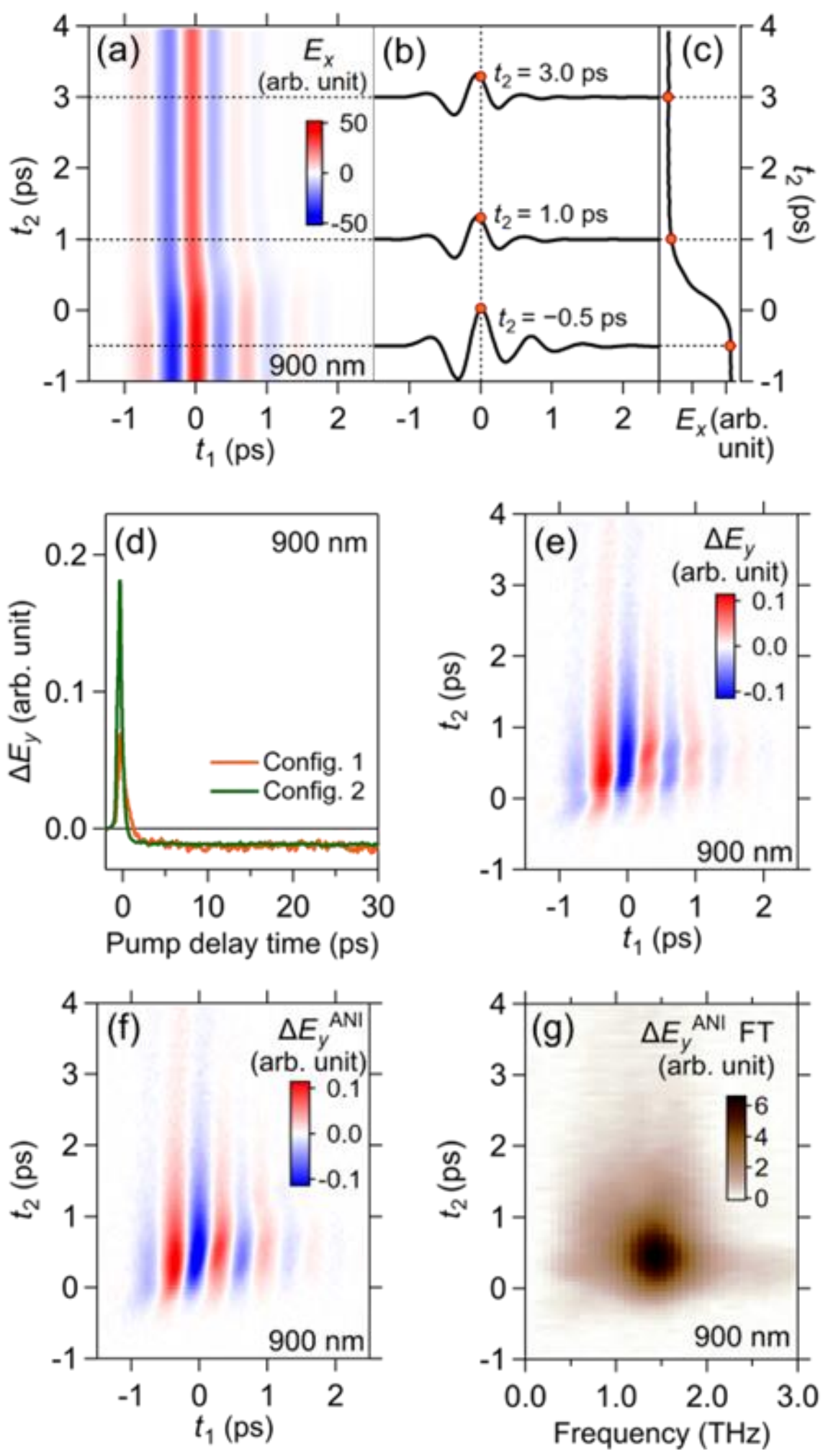

FIG. 2. (a) 2D plot of the transmitted THz probe field $E_x$ as a function of probe delay $t_1$ and pump delay $t_2$ for a 900 nm pump. (b) Waveforms of $E_x(t_1)$ at fixed pump delays. (c) Vertical profile of $E_x(t_2)$ at a fixed probe delay $t_1$=0 ps. (d) Dynamics of light-induced $y$-component $\Delta E_y$ as a function of the pump delay $t_2$ with a fixed probe delay. The orange and green curves correspond to Configuration 1 and 2, respectively. (e) 2D plot of $\Delta E_y(t_1, t_2)$. (f) 2D plot of anisotropic conductivity signal $\Delta E_y^{\mathrm{ANI}}(t_1, t_2)$. (g) Fourier transform of (f) at each delay $t_2$.

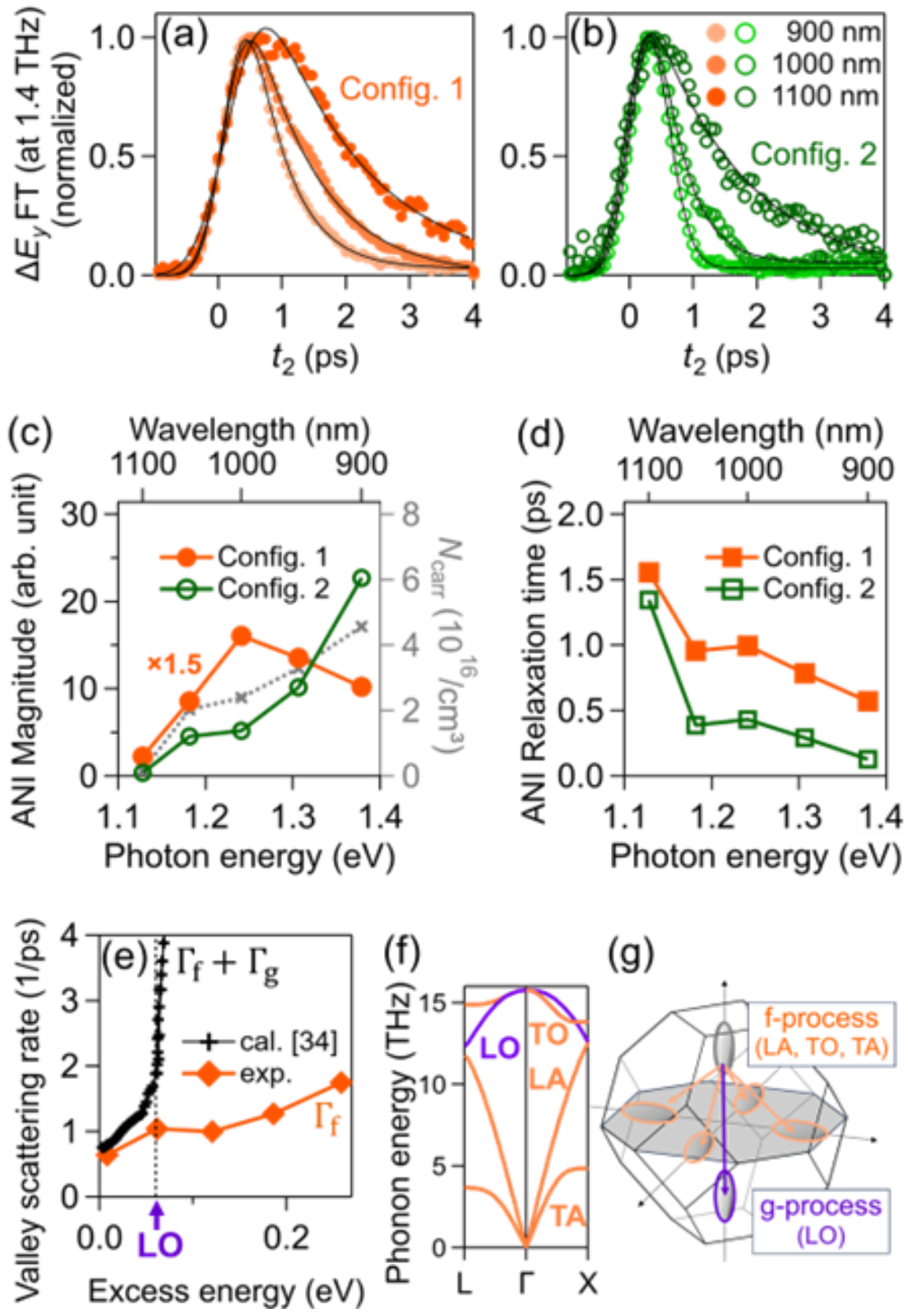

FIG. 3. (a)(b)$\Delta E_y$ at the probe frequency $\omega_1/2\pi$=1.4 THz as a function of pump delay $t_2$ in Configs. 1 and 2, respectively, for various pump wavelengths. The black solid curves are the fitting results. (c) Magnitude of the anisotropic conductivity (left axis) for CB and VB as a function of the pump photon energy. The right axis shows the photoexcited carrier density. (d) The relaxation time of CB valley polarization and VB momentum asymmetry as a function of the pump photon energy. (e) The scattering rate vs the excess energy from the band gap for the fitting results of this experiment and the previous calculation [34]. (f) The phonon dispersion relation in Si. (g) Schematics of intervalley scattering by phonons with $g$ and $f$ processes.

## References

[1] T. Ando, A. B. Fowler, and F. Stern, *Electronic properties of two-dimensional systems.* Rev. Mod. Phys. **54**, 437 (1982). https://doi.org/10.1103/RevModPhys.54.437

[2] J. Schaibley, H. Yu, G. Clark, P. Rivera, J. S. Ross, K. L. Seyler, W. Yao, and X. Xu, *Valleytronics in 2D materials.* Nat. Rev. Mater. **1**, 16055 (2016). https://doi.org/10.1038/natrevmats.2016.55

[3] K. F. Mak and J. Shan, *Photonics and optoelectronics of 2D semiconductor transition metal dichalcogenides.* Nat. Photon. **10**, 216 (2016). https://doi.org/10.1038/nphoton.2015.282

[4] A. Rycerz, J. Tworzydło, and C. Beenakker, *Valley filter and valley valve in graphene.* Nat. Phys. **3**, 172 (2007). https://doi.org/10.1038/nphys547

[5] D. Xiao, W. Yao and Q. Niu, *Valley-Contrasting Physics in Graphene: Magnetic Moment and Topological Transport.* Phys. Rev. Lett. **99**, 236809 (2007). https://doi.org/10.1103/PhysRevLett.99.236809

[6] D. Xiao, Gui-Bin Liu, W.. Feng, X. Xu, and W. Yao, *Coupled Spin and Valley Physics in Monolayers of MoS2 and Other Group-VI Dichalcogenides.* Phys. Rev. Lett. **108**, 196802 (2012). https://doi.org/10.1103/PhysRevLett.108.196802

[7] K. F. Mak, K. He, J. Shan, and T. F. Heinz, *Control of valley polarization in monolayer MoS2 by optical helicity.* Nat. Nanotech. **7**, 494 (2012). https://doi.org/10.1038/nnano.2012.96

[8] K. F. Mak, K. L. McGill, J. Park, and P. L. McEuen, *The valley Hall effect in MoS2 transistors.* Science **344**, 1489 (2014). https://doi.org/10.1126/science.1250140

[9] F. J. Ohkawa and Y. Uemura, *Theory of Valley Splitting in an N-Channel (100) Inversion Layer of Si II. Electric Break Through.* J. Phys. Soc. Jpn. **43**, 917 (1977). https://doi.org/10.1143/JPSJ.43.917

[10] L. J. Sham, S. J. Allen, Jr., A. Kamgar, D. C. Tsui, *Valley-Valley Splitting in Inversion Layers on a High-Index Surface of Si.* Phys. Rev. Lett. **40**, 472 (1978). https://doi.org/10.1103/PhysRevLett.40.472

[11] S. Takagi, J. L. Hoyt. J. J. Welser, and J. F. Gibbons, *Comparative study of phonon-limited mobility of two-dimensional electrons in strained and unstrained Si metal–oxide–semiconductor field-effect transistors.* J. Appl. Phys. **80**, 1567 (1996). https://doi.org/10.1063/1.362953

[12] S. E. Thompson *et al.*, *A 90-nm logic technology featuring strained-silicon. IEEE Trans.* Electron Devices **51**, 1790 (2004). https://doi.org/10.1109/TED.2004.836648

[13] S. Goswami, K. A. Slinker, M. Friesen, L. M. McGuire, J. L. Truitt, C. Tahan, L. J. Klein, J. O. Chu, P. M. Mooney, D. W. van der Weide, Robert Joynt, S. N. Coppersmith, and Mark A. Eriksson, *Controllable valley splitting in silicon quantum devices*. Nat. Phys. **3**, 41 (2007). https://www.nature.com/articles/nphys475

[14] B. Koiller, X. Hu, and S. Das Sarma, *Exchange in Silicon-Based Quantum Computer Architecture.* Phys. Rev. Lett. **88**, 027903 (2001). https://doi.org/10.1103/PhysRevLett.88.027903
[15] Y. P. Shkolnikov, E. P. De Poortere, E. Tutuc, and M. Shayegan, *Valley Splitting of AlAs Two-Dimensional Electrons in a Perpendicular Magnetic Field.* Phys. Rev. Lett. **89**, 226805 (2002). https://doi.org/10.1103/PhysRevLett.89.226805
[16] O. Gunawan, Y. P. Shkolnikov, K. Vakili, T. Gokmen, E. P. De Poortere, and M. Shayegan, *Valley Susceptibility of an Interacting Two-Dimensional Electron System.* Phys. Rev. Lett. **97**, 186404 (2006). https://doi.org/10.1103/PhysRevLett.97.186404
[17] J. Isberg, M. Gabrysch, J. Hammersberg, S. Majdi, K. K. Kovi, and D. J. Twitchen, *Generation, transport and detection of valley-polarized electrons in diamond.* Nat. Mater. **12**, 760 (2013). https://doi.org/10.1038/nmat3694
[18] M. Kozák, T. Otobe, M. Zukerstein, F. Trojánek, and P. Malý, *Anisotropy and polarization dependence of multiphoton charge carrier generation rate in diamond.* Phys. Rev. B **99**, 104305 (2019). https://doi.org/10.1103/PhysRevB.99.104305
[19] L. D. Laude, F. H. Pollak, and M. Cardona, *Effects of Uniaxial Stress on the Indirect Exciton Spectrum of Silicon.* Phys. Rev. B **3**, 2623 (1971). https://doi.org/10.1103/PhysRevB.3.2623
[20] A. A. Kaplyanskii and N. S. Sokolov, *Selective optical valley pumping in silicon and germanium.* Sol. Stat. Comm. **20**, 27 (1976). https://doi.org/10.1016/0038-1098(76)91691-4
[21] I. Akimoto and N. Naka, *Two optical routes of cold carrier injection in silicon revealed by time-resolved excitation spectroscopy.* Appl. Phys. Express **10**, 061301 (2017). https://doi.org/10.7567/APEX.10.061301
[22] R. N. Dexter, B. Lax, A. F. Kip, and G. Dresselhaus, *Effective masses of electrons in silicon.* Phys. Rev. **96**, 222 (1954). https://doi.org/10.1103/PhysRev.96.222
[23] T. Ichibayashi, S. Tanaka, J. Kanasaki, K. Tanimura and T. Fauster, *Ultrafast relaxation of highly excited hot electrons in Si: Roles of the $L-X$ intervalley scattering*. Phys. Rev. B **84**, 235210 (2011). http://dx.doi.org/10.1103/PhysRevB.84.235210
[24] J. Kanasaki, H. Tanimura, K. Tanimura, P. Ries, W. Heckel, K. Biedermann, and T. Fauster, *Ultrafast dynamics in photoexcited valence-band states of Si studied by time- and angle-resolved photoemission spectroscopy of bulk direct transitions.* Phys. Rev. B **97**, 035201 (2018). https://doi.org/10.1103/PhysRevB.97.035201
[25] H. Tanimura, J. Kanasaki, K. Tanimura, J. Sjakste, and N. Vast, *Ultrafast relaxation dynamics of highly excited hot electrons in silicon.* Phys. Rev. B **100**, 035201 (2019). https://doi.org/10.1103/PhysRevB.100.035201
[26] A. J. Sabbah and D. M. Riffe, *Femtosecond pump-probe reflectivity study of silicon carrier*

*dynamics.* Phys. Rev. B **66**, 165217 (2002). https://doi.org/10.1103/PhysRevB.66.165217
[27] M. Hase, M. Kitajima, A. M. Constantinescu, and H. Petek, *The birth of a quasiparticle in silicon observed in time–frequency space.* Nature **426**, 51 (2003). https://doi.org/10.1038/nature02044
[28] K. Kato, A. Ishizawa, K. Oguri, K. Tateno, T. Tawara, H. Gotoh, M. Kitajima, and H. Nakano, *Anisotropy in Ultrafast Carrier and Phonon Dynamics in p-Type Heavily Doped Si.* Jpn. J. Appl. Phys. **48**,100205 (2009). http://dx.doi.org/10.1143/JJAP.48.100205
[29] K. Ishioka, A. Rustagi, U. Hofer, H. Petek, and C. J. Stanton, *Intrinsic coherent acoustic phonons in the indirect band gap semiconductors Si and GaP.* Phys. Rev. B **95**, 035205 (2017). https://doi.org/10.1103/PhysRevB.95.035205
[30] S. K. Cushing, A. Lee, I. J. Porter, L. M. Carneiro, H.-T. Chang, M. Zürch, and S. R. Leone, *Differentiating Photoexcited Carrier and Phonon Dynamics in the Δ, L, and Γ Valleys of Si(100) with Transient Extreme Ultraviolet Spectroscopy.* J. Phys. Chem. C **123**, 3343 (2019). http://dx.doi.org/10.1021/acs.jpcc.8b10887
[31] M. Wörle, A. W. Holleitner, R. Kienberger, and H. Iglev, *Ultrafast hot-carrier relaxation in silicon monitored by phase-resolved transient absorption spectroscopy.* Phys. Rev. B **104**, L041201 (2021). https://doi.org/10.1103/PhysRevB.104.L041201
[32] C. Jacoboni and L. Reggiani, *The Monte Carlo method for the solution of charge transport in semiconductors with applications to covalent materials*. Rev. Mod. Phys. *55*, 645 (1983). https://doi.org/10.1103/RevModPhys.55.645
[33] T. Iizuka and M. Fukuma, *Carrier transport simulator for silicon based on carrier distribution function evolutions.* Solid-State Electronics **33**, 27 (1990). https://doi.org/10.1016/0038-1101(90)90005-Y
[34] J. Ma and S. Sinha, *Thermoelectric properties of highly doped n-type polysilicon inverse opals.* J. Appl. Phys. **112**, 073719 (2012). https://doi.org/10.1063/1.4758382
[35] E. Witkoske, X. Wang, M. Lundstrom, V. Askarpour, and J. Maassen, *Thermoelectric band engineering: The role of carrier scattering.* J. Appl. Phys. **122**, 175102 (2017). https://doi.org/10.1063/1.4994696
[36] Q.-L. Yang, W. Li, Z. Wang, F.-long Ning, and J.-W. Luo, *Uncovering the important role of transverse acoustic phonons in the carrier-phonon scattering in silicon.* Phys. Rev. B **109**, 125203 (2024). https://doi.org/10.1103/PhysRevB.109.125203
[37] D. G. Cooke, A. N. MacDonald, A. Hryciw, J. Wang, Q. Li, A. Meldrum, and F. A. Hegmann, *Transient terahertz conductivity in photoexcited silicon nanocrystal films.* Phys. Rev. B **73**, 193311 (2006). https://doi.org/10.1103/PhysRevB.73.193311
[38] L. Fekete, P. Kužel, H. Němec, F. Kadlec, A. Dejneka, J. Stuchlík, and A. Fejfar, *Ultrafast carrier dynamics in microcrystalline silicon probed by time-resolved terahertz spectroscopy.*

Phys. Rev. B **79**, 115306 (2009). https://doi.org/10.1103/PhysRevB.79.115306

[39] T. Suzuki and R. Shimano, *Time-Resolved Formation of Excitons and Electron-Hole Droplets in Si Studied Using Terahertz Spectroscopy.* Phys. Rev. Lett. **103**, 057401 (2009). https://doi.org/10.1103/PhysRevLett.103.057401

[40] T. Suzuki and R. Shimano, *Cooling dynamics of photoexcited carriers in Si studied using optical pump and terahertz probe spectroscopy.* Phys. Rev. B **83**, 085207 (2011). http://dx.doi.org/10.1103/PhysRevB.83.085207

[41] R. Ulbricht, E. Hendry, J. Shan, T. F. Heinz, and M. Bonn, *Carrier dynamics in semiconductors studied with time-resolved terahertz spectroscopy.* Rev. Mod. Phys. **83**, 543 (2011). https://doi.org/10.1103/RevModPhys.83.543

[42] T. Suzuki and R. Shimano, *Exciton Mott Transition in Si Revealed by Terahertz Spectroscopy.* Phys. Rev. Lett. **109**, 046402 (2012). https://doi.org/10.1103/PhysRevLett.109.046402

[43] T. Terashige, H. Yada, Y. Matsui, T. Miyamoto, N. Kida, and H. Okamoto, *Temperature and carrier-density dependence of electron-hole scattering in silicon investigated by optical-pump terahertz-probe spectroscopy.* Phys. Rev. B **91**, 241201(R) (2015). https://doi.org/10.1103/PhysRevB.91.241201

[44] S. Revuelta and E. Cánovas, *Contactless determination and parametrization of charge-carrier mobility in silicon as a function of injection level and temperature using time-resolved terahertz spectroscopy.* Phys. Rev. B **107**, 085204 (2023). https://doi.org/10.1103/PhysRevB.107.085204

[45] T. Moriyasu, M. Tani, H. Kitahara, T. Furuya, J. Afalla, T. Kohmoto, D. Koide, H. Sato, and M. Kumakura, *Photocarrier dynamics in thick Si film studied by optical pump-terahertz probe spectroscopy.* Optics Comm. **554**, 130139 (2024). https://doi.org/10.1016/j.optcom.2023.130139

[46] T. Fujimoto, T. Kurihara, Y. Murotani, T. Tamaya, N. Kanda, C. Kim, J. Yoshinobu, H. Akiyama, T. Kato, and R. Matsunaga, *Observation of Terahertz Spin Hall Conductivity Spectrum in GaAs with Optical Spin Injection.* Phys. Rev. Lett. **132**, 016301 (2024). https://doi.org/10.1103/PhysRevLett.132.016301

[47] See Supplemental Material at [link] for our experimental setup and analysis, which includes Refs. [48-62]

[48] C.-H. Lu, Y.-J. Tsou, H.-Y. Chen, B.-H. Chen, Y.-C. Cheng, S.-D. Yang, M.-C. Chen, C.-C. Hsu, and A. H. Kung, *Generation of intense supercontinuum in condensed media.* Optica **1**, 400 (2014). https://doi.org/10.1364/OPTICA.1.000400

[49] N. Kanda, N. Ishii, J. Itatani, and R. Matsunaga, *Optical parametric amplification of phase-stable terahertz-to-mid-infrared pulses studied in the time domain.* Optics Express **29**, 3479 (2021). https://doi.org/10.1364/OE.413200
[50] J. T. Kindt and C. A. Schmuttenmaer, *Theory for determination of the low-frequency time-dependent response function in liquids using time-resolved terahertz pulse spectroscopy.* J. Chem. Phys. **110**, 8589 (1999). https://doi.org/10.1063/1.478766
[51] M. A. Green, *Self-consistent optical parameters of intrinsic silicon at 300 K including temperature coefficients*. Solar Energy Materials and Solar Cells **92**, 1305 (2008). https://doi.org/10.1016/j.solmat.2008.06.009
[52] A. A. Bakun, B. P. Zakharchenya, A. A. Rogachev, M. N. Tkachuk, and V. G. Fleisher, *Observation of a surface photocurrent caused by optical orientation of electrons in a semiconductor.* JETP Lett. **40**, 1293 (1984). http://www.jetpletters.ru/ps/1262/article_19087.shtml
[53] K. J. Willis, S. C. Hagness, and I. Knezevic, *A generalized Drude model for doped silicon at terahertz frequencies derived from microscopic transport simulation*. Appl. Phys. Lett. **102**, 122113 (2013). https://doi.org/10.1063/1.4798658
[54] M. C. Beard, G. M. Turner, and C. A. Schmuttenmaer, *Transient photoconductivity in GaAs as measured by time-resolved terahertz spectroscopy.* Phys. Rev. B **62**, 15764 (2000). https://doi.org/10.1103/PhysRevB.62.15764
[55] N. V. Smith, *Classical generalization of the Drude formula for the optical conductivity.* Phys. Rev. B **64**, 155106 (2001). https://doi.org/10.1103/PhysRevB.64.155106
[56] N. O. Lipari and A. Baldereschi, *Energy Levels of Indirect Excitons in Semiconductors with Degenerate Bands.* Phys. Rev. B **3**, 2497 (1971). https://doi.org/10.1103/PhysRevB.3.2497
[57] K. L. Shaklee and R. E. Nahory, *Valley-orbit splitting of free excitons? The absorption edge of Si.* Phys. Rev. Lett. **24**, 942 (1970). https://doi.org/10.1103/PhysRevLett.24.942
[58] M. N. Saha, *On a physical theory of stellar spectra.* Proc. R. Soc. Lond. A **99**, 135 (1921). https://doi.org/10.1098/rspa.1921.0029
[59] T. Boggess, K. Bohnert, K. Mansour, S. Moss, I. Boyd, and A. Smirl, *Simultaneous measurement of the two-photon coefficient and free-carrier cross section above the bandgap of crystalline silicon* IEEE Journal of Quantum Electronics 22, 360 (1986). https://ieeexplore.ieee.org/document/1072964
[60] H. Němec, F. Kadlec, and P. Kužel, *Methodology of an optical pump-terahertz probe experiment: An analytical frequency-domain approach.* J. Chem. Phys. **117**, 8454 (2002). https://doi.org/10.1063/1.1512648

[61] T. Fujimoto, Y. Murotani, T. Tamaya, T. Kurihara, N. Kanda, C. Kim, J. Yoshinobu, H. Akiyama, T. Kato, and R. Matsunaga, *Transverse Current Generation by Circularly Polarized Light in Terahertz-Biased Semiconductor.* arXiv:2411.00528. https://arxiv.org/abs/2411.00528
[62] B. N. Brockhouse, *Lattice Vibrations in Silicon and Germanium*. Phys. Rev. Lett. **2**, 256 (1959). https://doi.org/10.1103/PhysRevLett.2.256
[63] M. Shalaby, M. Peccianti, Y. Ozturk, M. Clerici, I. Al-Naib, L. Razzari, T. Ozaki, A. Mazhorova, M. Skorobogatiy, and R. Morandotti, *Terahertz Faraday rotation in a magnetic liquid: High magneto-optical figure of merit and broadband operation in a ferrofluid.* Appl. Phys. Lett. **100**, 241107 (2012). https://doi.org/10.1063/1.4729132
[64] Y. Lubashevsky, LiDong Pan, T. Kirzhner, G. Koren, and N. P. Armitage, *Optical Birefringence and Dichroism of Cuprate Superconductors in the THz Regime*. Phys. Rev. Lett. **112**, 147001 (2014). https://doi.org/10.1103/PhysRevLett.112.147001
[65] F. Cerdeira, T. A. Fjeldly, and M. Cardona, *Effect of Free Carriers on Zone-Center Vibrational Modes in Heavily Doped $p$-type Si. II. Optical Modes*. Phys. Rev. B **8**, 4734 (1973). https://doi.org/10.1103/PhysRevB.8.4734
[66] M. A. Kanehisa, R. F. Wallis, and M. Balkanski, *Interband electronic Raman scattering in $p$-silicon.* Phys. Rev. B **25**, 7619 (1982). https://doi.org/10.1103/PhysRevB.25.7619
[67] R. Scholz, T. Pfeifer, and H. Kurz, *Density-matrix theory of coherent phonon oscillations in germanium.* Phys. Rev. B **47**, 16229 (1993). https://doi.org/10.1103/PhysRevB.47.16229
[68] M. V. Dolguikh and R. E. Peale, *Anisotropic optical phonon scattering of holes in cubic semiconductors*. J. Appl. Phys. **101**, 113716 (2007). https://doi.org/10.1063/1.2745222
[69] J. R. Chelikowsky and M. L. Cohen, *Electronic band structure od silicon.* Phys. Rev. B **10**, 5095 (1974). https://doi.org/10.1103/PhysRevB.10.5095
[70] S. D. Conte, F. Bottegoni, E. A. A. Pogna, D. D. Fazio, S. Ambrogio, I. Bargigia, C. D'Andrea, A. Lombardo, M. Bruna, F. Ciccacci, A. C. Ferrari, G. Cerullo, and M. Finazzi, *Ultrafast valley relaxation dynamics in monolayer MoS2 probed by nonequilibrium optical techniques*. Phys. Rev. B **92**, 235425 (2015). http://dx.doi.org/10.1103/PhysRevB.92.235425
[71] H. W. Streitwolf, *Intervalley scattering selection rules for Si and Ge*. phys. stat. sol. **37**, K47 (1970). https://onlinelibrary.wiley.com/doi/epdf/10.1002/pssb.19700370165